\documentclass[aps,longbibliography,
  twocolumn,amsmath,
  amssymb,nofootinbib, superscriptaddress, preprintnumbers]{revtex4-1}

\usepackage{graphics}      
\usepackage{graphicx}      
\usepackage{longtable}     
\usepackage{url}           
\usepackage{bm}            
\usepackage{xcolor}
\usepackage{mathrsfs}
\usepackage[colorlinks,allcolors=blue]{hyperref}
\usepackage{blindtext}
\usepackage{hyperref}
\usepackage{float}
\usepackage{bbold}
\usepackage{lipsum}
\usepackage[normalem]{ulem}
\usepackage{orcidlink}
\usepackage{academicons}
\usepackage{xcolor}
\usepackage{subfigure}
\usepackage{booktabs}
\usepackage{hyperref}

\newcommand{\orcid}[1]{\href{https://orcid.org/#1}{\textcolor[HTML]{A6CE39}{\aiOrcid}}}

\begin{document}
\preprint{MPP-2024-11}

\title{Using Bayesian Inference to Distinguish Neutrino Flavor Conversion Scenarios via a Prospective Supernova Neutrino Signal}


\newcommand*{\MPP}{\textit{\small{Max-Planck-Institut f\"ur Physik (Werner-Heisenberg-Institut), Boltzmannstr. 8, 85748 Garching, Germany}}}

\author{Sajad Abbar \orcidlink{0000-0001-8276-997X}   } 
\affiliation{\MPP}

\author{Maria Cristina Volpe}
\affiliation{Astro-Particule et Cosmologie (APC), CNRS UMR 7164, Universite Denis Diderot, 10, rue Alice Domon et Leonie Duquet, 75205 Paris Cedex 13, France }


\begin{abstract}

The upcoming galactic core-collapse supernova  is expected to produce a considerable number of  neutrino events within terrestrial detectors. 
By using  Bayesian inference techniques, we address the feasibility of 
 distinguishing among various neutrino flavor conversion   scenarios in the supernova environment, using such a neutrino signal.
 In addition to the conventional MSW, we explore several more sophisticated flavor conversion scenarios, such as spectral swapping, fast flavor conversions, flavor equipartition caused by non-standard neutrino interactions, magnetically-induced flavor equilibration, and flavor equilibrium resulting from slow flavor conversions. 
 Our analysis demonstrates that with a sufficiently large number of neutrino events during the supernova accretion phase (exceeding several hundreds), there exists a good probability of distinguishing among feasible neutrino flavor conversion scenarios in the supernova environment.

 \end{abstract}

\maketitle

\section{Introduction}

Neutrinos originating from core-collapse supernovae (CCSNe) present an unprecedented opportunity for exploring the intricacies of astroparticle physics. 
Understanding the behavior and characteristics of these neutrinos can significantly contribute to the broader comprehension of  neutrino physics in extreme conditions and offers invaluable insights into the physics governing the CCSNe~\cite{Horiuchi:2018ofe, Raffelt:2007nv, Janka:2017vcp, volpe2023neutrinos}.

Since the detection of Supernova 1987A (SN 1987A) and its corresponding neutrino signal, there has been a substantial body of research focusing on studying a galactic SN neutrino signals for various purposes. These studies have  focused on a number of issues,
such as developing insight into the SN explosion mechanism~\cite{Loredo:2001rx, Pagliaroli:2008ur},
understanding neutrino spectral parameters~\cite{Harada:2023elm, GalloRosso:2018ugl, GalloRosso:2017mdz}, analyzing SN remnant features~\cite{Suwa:2022vhp, GalloRosso:2017hbp}, and exploring non-standard physics in extreme conditions~\cite{Kolb:1987qy, Shalgar:2019rqe, deGouvea:2022dtw}. Specifically,   Refs.~\cite{Hyper-Kamiokande:2021frf,Olsen:2022pkn,Saez:2024ayk} have recently employed  Bayesian techniques  to differentiate between various SN models by analyzing a future SN neutrino signal.

One of the most captivating aspects of SN neutrinos is that they are thought to undergo a phenomenon known as \emph{collective} neutrino oscillations~\cite{pantaleone:1992eq, sigl1993general, Pastor:2002we,duan:2006an, duan:2006jv, duan:2010bg, Mirizzi:2015eza,volpe2023neutrinos}. This intriguing behavior arises from 
an intricate interplay between propagating neutrinos and the dense background neutrino gas, where the coherent forward scatterings of neutrinos play a major role.
What makes this phenomenon particularly fascinating is its nonlinear and collective nature.

The phenomenon of neutrino flavor conversion (FC) within the  SN environment displays a remarkable richness. It has been established that varying the physics conditions in the SN environment can give rise to different FC scenarios. While the conventional Mikheev-Smirnov-Wolfenstein (MSW) effect arises from a resonant behavior resulting from the interplay between matter and vacuum Hamiltonian terms, the SN environment introduces more intricate FC scenarios for neutrinos. 

Initially, the understanding of neutrino FCs in the  SN environment was mostly centered around the so-called phenomenon of slow modes (driven by the neutrino vacuum frequency), and specifically
the well-established phenomenon of spectral swapping~\cite{duan:2006an, duan:2006jv, duan:2010bg, dasgupta:2009mg, Galais:2011gh, Dasgupta:2021gfs}. This phenomenon involves the interchange of spectra between $\nu_e$($\bar\nu_e$) and $\nu_x$($\bar\nu_x$) across a range of neutrino energies. 

However, further exploration revealed that the actual FC scenarios within more realistic supernova models can be considerably diverse.
For example, studies have demonstrated the existence of fast flavor conversions (FFCs) when the angular distributions of $\nu_e$ and $\bar\nu_e$ exhibit significant enough differences, so that they can cross each other~\cite{sawyer2005speed, sawyer2016neutrino, Morinaga:2021vmc} (see~\cite{volpe2023neutrinos} for a review on flavor mechanisms).
 
Additionally, in the presence of strong magnetic fields, (partial) flavor equilibration between neutrinos and antineutrinos may occur due to the presence of the neutrino magnetic moment~\cite{Abbar:2020ggq, Sasaki:2021bvu, Kharlanov:2020cti}.   Another interesting FC scenario occurs when neutrinos can experience 
non-standard self interactions (NSSI). Such NSSI introduce yet another facet of flavor equilibration among various neutrino (antineutrino) flavors~\cite{Abbar:2022jdm, Stapleford:2016jgz}.

In this paper, we utilize Bayesian inference techniques to discern between various neutrino FC  scenarios by analyzing a future  SN neutrino signal. Our specific focus lies within the SN accretion phase, characterized by distinctive energy spectra among different neutrino species. We demonstrate that a sufficiently large number of neutrino events significantly enhances the likelihood of distinguishing between different FC scenarios in the SN environment.

The paper is structured as follows.
 In Sec.~\ref{sec:spec}, we describe the SN neutrino spectra.
  We then discuss the neutrino FC scenarios  in Sec.~\ref{sec:FC}. 
  Sec.~\ref{sec:cross} 
   introduces the neutrino detection channels we consider in this study.
  Before concluding in Sec.~\ref{sec:dis}, we present the results of our Bayesian analysis of a prospective SN neutrino signal in Sec.~\ref{sec:BF}.

 \section{SN neutrino signal}\label{sec:spec}
 
To ensure the maximum impact of the FC on neutrino energy spectra, we strategically select the  SN accretion phase, anticipating the greatest disparity between different neutrino spectra. 
As a first step, we make the simplifying hypothesis that  the neutrino  spectral parameters are stationary during the emission of the neutrino signal.
This assumption can be justified on the basis of two key factors.
Firstly, during the SN accretion phase, the neutrino spectral parameters can reasonably be considered constant, at least for several tens of milliseconds, probably allowing for the accumulation of a sufficient number of events. Additionally, our methodology allows us to determinine the minimum time interval for which the spectral parameters should remain constant to obtain any conclusive result.

In scenarios where neutrino FC is not present, the time-integrated neutrino energy-differential number flux   of a specific neutrino species,  $\nu_\beta$, at a distance $D$ from the SN can be written as~\cite{Mirizzi:2015eza},
\begin{equation}
\mathcal{F_{\nu_\beta}}(E_\nu) = \frac{\mathcal{E}_{\nu_\beta}}{4 \pi D^2 \langle E_{\nu_\beta} \rangle} f_{\nu_\beta}(E_\nu)
\end{equation}
with
\begin{equation}
f_{\nu_\beta}(E_\nu) = \frac{1}{T_{\nu_\beta} \Gamma(1+\alpha_{\nu_\beta})} \bigg( \frac{E_\nu}{T_{\nu_\beta}} \bigg)^{\alpha_{\nu_\beta}}  \exp(-E_\nu/T_{\nu_\beta})
\end{equation}
being the normalized $\nu_\beta$ spectrum, where $E_\nu$ is the neutrino energy. Here, $\alpha_{\nu_\beta}$ and $\langle E_{\nu_\beta} \rangle$ 
are the pinching parameter and the average neutrino energy, 
 the parameters which describe  the normalized spectrum and $T_{\nu_\beta} = \langle E_{\nu_\beta} \rangle/(1+\alpha_{\nu_\beta})$. In addition, $\mathcal{E}_{\nu_\beta}$ is the  time-integrated neutrino luminosity for the time interval of interest, i.e.,  $\mathcal{E}_{\nu_\beta} = \int_{t}^{t+\Delta t} \mathrm{d}t L_{\nu_\beta}(t)$, with $L_{\nu_\beta}$ being the neutrino luminosity.

\begin{table}[b]
\centering
\caption{The  neutrino
energy spectra we consider in this study, relevant to the SN accretion phase. 
While the LSD spectrum shows significant variations among neutrino species, in the SSD one the  spectral variations are less prominent~\cite{Serpico:2011ir,Fischer:2009af}.}
\begin{tabular}[t]{|l|}
\hline
\textcolor{black}{ \quad \quad \quad \quad \quad \quad \textbf{{ LSD}}   } \\
\hline
$\mathcal{E}_{\nu_e}: \mathcal{E}_{\bar\nu_e}: \mathcal{E}_{\nu_x} = 1:1:0.33$ \\
$\langle E_{\nu_e} \rangle: \langle E_{\bar\nu_e} \rangle: \langle E_{\nu_x} \rangle = 9:12:16.5$ \\
${\alpha}_{\nu_e}: {\alpha}_{\bar\nu_e}: {\alpha}_{\nu_x} = 3.2:4.5:2.3$ \\
\hline
\textcolor{black}{ \quad \ \quad \quad \quad \quad \quad  \textbf{{SSD}}  }\\
\hline
$\mathcal{E}_{\nu_e}: \mathcal{E}_{\bar\nu_e}: \mathcal{E}_{\nu_x} = 1:1:0.75$ \\
$\langle E_{\nu_e} \rangle: \langle E_{\bar\nu_e} \rangle: \langle E_{\nu_x} \rangle = 9:12:14$ \\
${\alpha}_{\nu_e}: {\alpha}_{\bar\nu_e}: {\alpha}_{\nu_x} = 3.26:4.76:2.26$ \\
\hline
\end{tabular}
\label{tab:spec}
\end{table}%

In the following, we analyze two distinctive neutrino energy spectra relevant to the SN accretion phase, as described in Table.~\ref{tab:spec}: one with large spectral difference (LSD), showing significant variations among neutrino species, and another with small spectral difference (SSD), where spectral variations are less prominent.
Here, we make the assumption that $\nu_x$ and
$\bar{\nu}_x$ have identical spectral shapes.
Also note that the average energies are in MeV, and that there is no unit in front of  $\mathcal{E}$, given that we 
here consider the scenario where the number of observed neutrino event are fixed (see the next section for more details).
While the latter profile is more suitable for SN models with lighter progenitors, the former is more appropriate for models involving heavier ones~\cite{Serpico:2011ir, Fischer:2009af}.

\section{Neutrino flavor conversion scenarios}\label{sec:FC}

In this section, we briefly describe the FC mechanisms under consideration. It is important to highlight that apart from the conventional MSW mechanism, all the FC scenarios we explore stem from coherent neutrino-neutrino scatterings (mean-field approximation). As a result, the total number of (anti)neutrinos within a designated energy/momentum range remains conserved.

\begin{figure*} [tb!]
\centering
\begin{center}
\includegraphics*[width=1.\textwidth, trim= 0 0 0 0, clip]{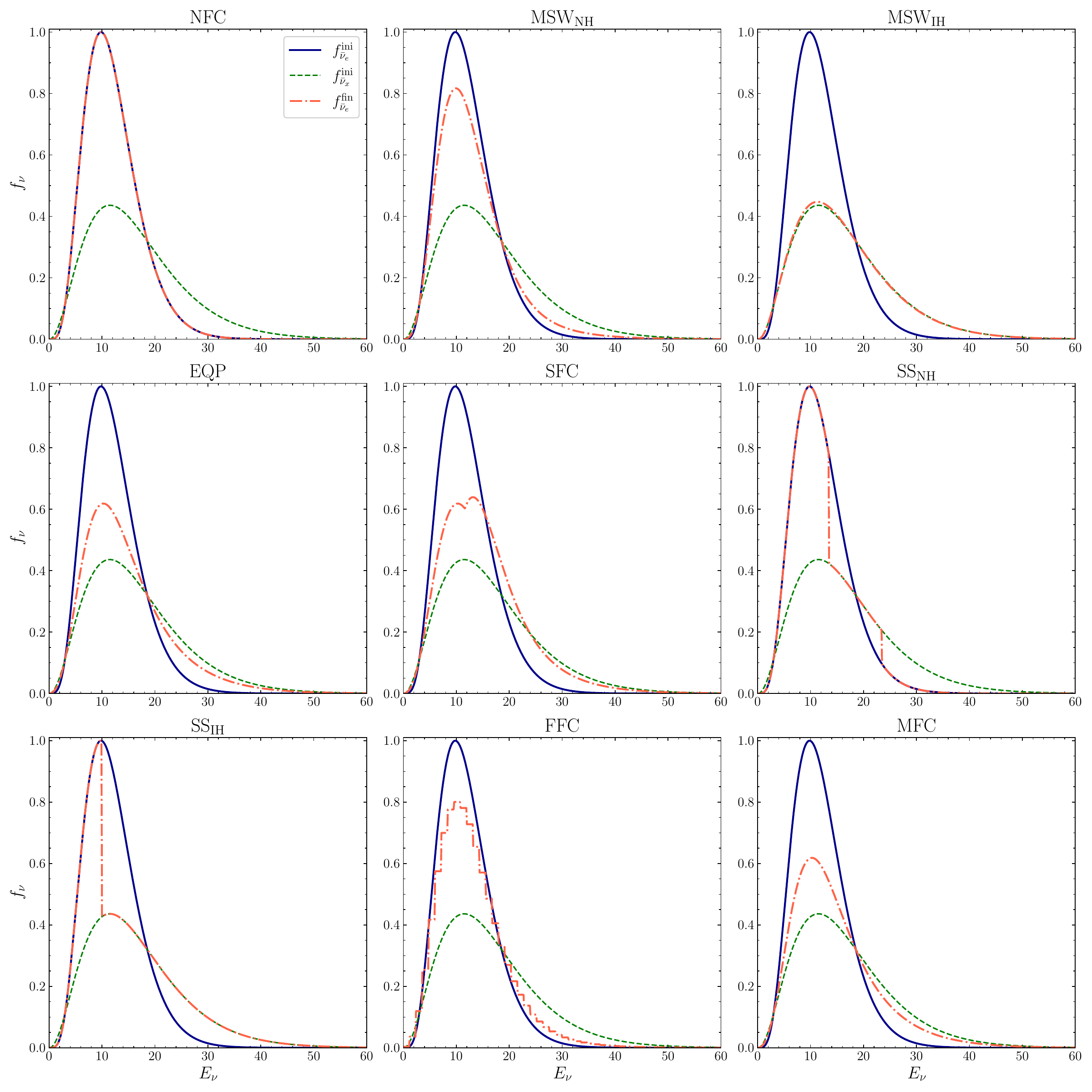}
\end{center}
\caption{
Initial spectra of $\bar\nu_e$ and $\bar\nu_x$ (derived from the LSD profile), as well as the resulting $\bar\nu_e$ spectrum post FC for different FC scenarios. Here, the initial spectra (LSD) are assumed to exist at the neutrinosphere. Then, we have applied the FC scenarios (defined in Sec.~\ref{sec:FC}) given in the title of each panel. }
\label{fig:spectra}
\end{figure*}

To provide the reader with an impression of the various FC scenarios, Fig.~\ref{fig:spectra} displays the initial spectra of $\bar\nu_e$ and $\bar\nu_x$ derived from the LSD profile, alongside the resulting $\bar\nu_e$ spectrum post the given  FC scenario.

\subsection{No flavor conversions (NFC)}
In our first example, we examine the FC scenario where the emitted neutrino fluxes within the  SN core precisely translates to larger distances without alteration. This scenario serves as a benchmark for comparison. Surprisingly, while seemingly idealistic, this scenario is not entirely implausible, 
given the fact we do not yet
know to which extent FC in the SN core influences the SN
neutrino fluxes.

\subsection{MSW effect}
The established MSW mechanism stands as one of the most extensively explored neutrino FC mechanisms. It occurs due to a resonant behavior generated by a cancellation in the diagonal term of the Hamiltonian, between the matter and vacuum  components. 
This effect is highly sensitive to the neutrino mass hierarchy. Vacuum neutrino oscillation experiments provide accurate knowledge of mass-squared differences. Due to matter effects in the Sun, we also know that $\Delta m_{21}^2>0$. However, the sign of the atmospheric mass splitting, $\Delta m_{31}^2$, is currently unknown since current measurement solely revealing its absolute value. Hence, we can have either a normal hierarchy (NH) with $\Delta m_{31}^2>0$, or an inverted hierarchy (IH) with $\Delta m_{31}^2<0$.
Assuming  a perfect adiabatic MSW, 
in NH 
one has~\cite{Dighe:1999bi},
\begin{equation}\label{eq:nh}
\begin{split}
\mathcal{F}_{\nu_e} &= \sin^2\theta_{13}\ \mathcal{F}^0_{\nu_e} +  \cos^2\theta_{13}\ \mathcal{F}^0_{\nu_x},\\
\mathcal{F}_{\bar\nu_e} &= \cos^2\theta_{12}\cos^2\theta_{13}\ \mathcal{F}^0_{\bar\nu_e} + (1- \cos^2\theta_{12}\cos^2\theta_{13})\ \mathcal{F}^0_{\bar\nu_x},\\
2\mathcal{F}_{\nu_x} &= \cos^2\theta_{13}\ \mathcal{F}^0_{\nu_e} +  (1+\sin^2\theta_{13})\ \mathcal{F}^0_{\nu_x},\\
2\mathcal{F}_{\bar\nu_x} &= (1-\cos^2\theta_{13}\cos^2\theta_{12})\ \mathcal{F}^0_{\bar\nu_e} +  (1+\cos^2\theta_{13}\cos^2\theta_{12})\ \mathcal{F}^0_{\bar\nu_x},\\
\end{split}
\end{equation}
while for IH, it is given by,
\begin{equation}\label{eq:ih}
\begin{split}
\mathcal{F}_{\nu_e} &= \cos^2\theta_{13}\sin^2\theta_{12}\ \mathcal{F}^0_{\nu_e} + (1- \sin^2\theta_{12} \cos^2\theta_{13})\ \mathcal{F}^0_{\nu_x},\\
\mathcal{F}_{\bar\nu_e} &= \sin^2\theta_{13}\ \mathcal{F}^0_{\bar\nu_e} + \cos^2\theta_{13}\ \mathcal{F}^0_{\bar\nu_x},\\
2\mathcal{F}_{\nu_x} &= (1-\cos^2\theta_{13}\sin^2\theta_{12})\ \mathcal{F}^0_{\nu_e} +  (1+\sin^2\theta_{12}\cos^2\theta_{13})\ \mathcal{F}^0_{\nu_x},\\
2\mathcal{F}_{\bar\nu_x} &= \cos^2\theta_{13}\ \mathcal{F}^0_{\bar\nu_e} +  (1+\sin^2\theta_{13})\ \mathcal{F}^0_{\bar\nu_x}.\\
\end{split}
\end{equation}
Here $\mathcal{F}^0_{\nu}$ denotes the neutrino fluxes at the neutrinosphere before
flavor conversion mechanisms produce spectral distortions.

\subsection{Spectral swapping (SS)}
The initial investigations into collective neutrino oscillations within the  SN environment focused on simplified symmetric models, such as the stationary spherically symmetric neutrino bulb model. These studies revealed a significant phenomenon: spectral swapping phenomenon
which involves the swapping of spectra between $\nu_e$($\bar\nu_e$) and $\nu_x$($\bar\nu_x$) over a range of neutrino energies, due to the collective oscillations of SN neutrinos~\cite{duan:2006an, duan:2006jv, duan:2010bg, dasgupta:2009mg, Dasgupta:2021gfs, volpe2023neutrinos}. Though this was first observed  in stationary spherically symmetric  models, in Ref.~\cite{Martin:2019dof} the authors showed that SS can also exist in multi-dimensional models provided that FCs do not occur very deep inside the SN core.

The spectral swap can occur around the neutrino energies where there is a zero crossing in the energy spectrum  defined as,
\begin{equation}
g_\omega = \frac{|\Delta m^2|}{2\omega^2} 
\begin{cases} 
& \mathcal{F}^0_{\nu_e} - \mathcal{F}^0_{\nu_x} \quad (\omega>0) \\
&\mathcal{F}^0_{\bar\nu_x} - \mathcal{F}^0_{\bar\nu_e} \quad (\omega<0) \\
\end{cases} 
\end{equation}
where $\omega =  \frac{|\Delta m^2|}{2E_\nu} $ and $|\Delta m^2|$ is the neutrino mass squared difference. 
Note that there can be a crossing (by default) when $\omega \rightarrow 0$, corresponding to $E_\nu \rightarrow \infty$, which if unstable
leads to an SS for the tail of the energy spectra.

The spectral crossings exhibit instability in the IH  (NH) scenarios when the crossing slope is positive (negative), resulting in a swap across a finite range of $\omega$. While zero crossings in $g_\omega$ serve as a criterion for instability concerning SS, determining the swapping width is not straightforward. In practice, there might exist even crossings that do not induce any spectral swapping.

In our computations, we prioritize practicality over ambition when estimating the width of the SS. Specifically, if the crossing at $\omega=0$ is unstable, we approximate $\Delta \omega \simeq 0.7$~km$^{-1}$. Alternatively, for each crossing, we assume that SS occurs within $\Delta E_\nu = \pm 7.5$~MeV. 
These assumptions align with the simulation results~\cite{duan:2006an, dasgupta:2009mg}. Furthermore, we have verified that the width of the crossings, if sufficiently large, does not impact the results presented in the subsequent section.

 \subsection{Slow flavor conversions (SFC)}
In the context of slow modes, in some symmetric models like the bulb model, it has been shown that some
flavor equilibration can be reached~\cite{Raffelt:2007yz}.
 Such equilibrium leads to neutrino flavor equipartition  up to the constraint
of neutrino electron lepton number conservation. 
The resulted neutrino spectra can then be found as,
\begin{equation}\label{Eq:equilibrium1}
    \begin{split}
        \mathcal{F}_{\nu_e}   & =  \mathcal{F}_{\rm{eq}} + \max(0, \mathcal{L} ), \\
        \mathcal{F}_{\bar\nu_e}  & =  \mathcal{F}_{\rm{eq}}  + \max(0,-\mathcal{L} ),\\
        \mathcal{F}_{\nu_x}   & =   \mathcal{F}_{\bar\nu_x} =  \mathcal{F}_{\rm{eq}},
    \end{split}
\end{equation}
where $\mathcal{F}_{\rm{eq}}$ is the equilibrium $\mathcal{F}$, defined by, 
    \begin{equation}
        \mathcal{F}_{\rm{eq}} =
        \begin{cases}
             \frac{1}{3}(\mathcal{F}^0_{\bar\nu_e}  + 2 \mathcal{F}^0_{\bar\nu_x} ),    & \text{if } \mathcal{L} > 0, \\
             \frac{1}{3}(\mathcal{F}^0_{\nu_e}  + 2 \mathcal{F}^0_{\nu_x} ),       & \text{if } \mathcal{L}  \leq 0.
        \end{cases}
    \end{equation} 
 where $\mathcal{L} = \mathcal{F}^0_{\nu_e}  - \mathcal{F}^0_{\bar\nu_e} $, provides information on the neutrino electron lepton number.
 For more details on the SFC prescription, we refer an interested reader to Ref.~\cite{Ehring:2023abs, Ehring:2023lcd}.

\subsection{Fast flavor conversions  (FFC)}
One of the most recent developments in the field of neutrino FC in the SN environment has been the discovery of FFCs~\cite{Sawyer:2005jk, Sawyer:2015dsa,Chakraborty:2016lct}.  
The assessment of the outcomes of FFCs has been extensively explored through localized dynamical simulations conducted within confined spaces using periodic boundary conditions~\cite{Bhattacharyya:2020dhu,Bhattacharyya:2020jpj,Wu:2021uvt,Richers:2021nbx,Zaizen:2021wwl,Richers:2021xtf,Bhattacharyya:2022eed,Grohs:2022fyq,Abbar:2021lmm,Richers:2022bkd,Zaizen:2022cik,Xiong:2023vcm}. Insights obtained from these investigations reveal a tendency toward kinematic decoherence during flavor conversions, leading to the emergence of quasistationary states. These stable states can be characterized by survival probabilities, governed by the conservation of neutrino lepton number, and have shown potential for accurate analytical modeling~\cite{Xiong:2023vcm}.

In this study, we take the analytical survival probability developed in Ref.~\cite{Xiong:2023vcm}, assuming an axisymmetric neutrino angular distributions.
In addition, we assumed the maximum entropy angular  distribution ~\cite{Cernohorsky:1994yg}, defined as,
\begin{equation}
\mathcal{G}_\nu(\mu) = \exp(\eta + a\mu),
\end{equation}
where  $\eta$ and $a$ are some arbitrary parameters and $\mu = \cos\theta_\nu$, with $\theta_\nu$ being the zenith angle of the neutrino velocity, and,
\begin{equation}
\mathcal{G}_\nu(\mu) =  \int_0^\infty \int_0^{2\pi} \frac{E_\nu^2 \mathrm{d} E_\nu \mathrm{d} \phi_\nu}{(2\pi)^3} \
        \mathcal{G}_{\nu}(\mathbf{p}),
\end{equation}
where $\mathcal{G}_{\nu}(\mathbf{p})$'s are the neutrino 
occupation numbers of different flavors.
Here,  $\phi_\nu$ is the  azimuthal angle of the neutrino velocity. 
If one is then provided with the flux factors of the energy-integrated as well as those of each energy bin, one can
find the outcome of FFC for each energy bin. 
Here, the flux factor is defined as,
\begin{equation}
    \mathbf{F}_\nu = \frac{I_1}{I_0}
\end{equation}
where $I$'s are the radial moments of the neutrino angular distributions, defined as,
\begin{equation}
I_n = \int_{-1}^{1} \mathrm{d}\mu\ \mu^n\ \int_0^\infty \int_0^{2\pi} \frac{E_\nu^2 \mathrm{d} E_\nu \mathrm{d} \phi_\nu}{(2\pi)^3} \
        \mathcal{G}_{\nu}(\mathbf{p}),
\end{equation}
for the energy-integrated moments, and as,
\begin{equation}
I_{n,i} = \frac{E_{\nu,i}^2 \Delta E_{\nu,i}}{(2\pi)^3}  \int_{-1}^{1} \mathrm{d}\mu\ \mu^n\ \int_0^{2\pi} \mathrm{d} \phi_\nu\
        \mathcal{G}_{\nu}(\mathbf{p}),
\end{equation}
when the corresponding quantities for each energy bin are required.
Here, $E_{\nu,i}$ and $\Delta E_{\nu,i}$  are the mean energy and the width of the i-th energy bin.
For the the energy-integrated flux factors, we take:
$\mathbf{F}_{\nu_e} = 0.5$, $\mathbf{F}_{\bar\nu_e} = 0.7$, and $\mathbf{F}_{\bar\nu_e} = 0.8$. Then for each energy bin we  adopted an assumption describing the relationship as $\mathbf{F}_{\nu,i} =\mathbf{F}_{\nu}\ (70 - E_{\nu,i})^2 /60^2$. Here, we have assumed that the spectra approach zero for $E_{\nu,i} \gtrsim 60$~MeV. 
Note that it is anticipated that the flux factor will decrease with increasing neutrino energy, and this reduction is expected to be nonlinear, due to the nonlinear scaling of the neutrino scattering cross-section (with the matter) with neutrino energy in the  SN environment. While our assumption regarding the dependence of the flux factor on energy is speculative, it aligns with both of these anticipated conditions. 
Otherwise, we regard this assumption as an initial exploration, considering it as the first stage of such a study. Further investigations may involve exploring alternative formulations for the energy dependence of the flux factor.

\subsection{Neutrino flavor equipartition (EQP)}

The existence of neutrino non-standard  self-interactions (NSSI) has been established as a mechanism capable of inducing equipartition between neutrinos and antineutrinos separately~\cite{Abbar:2022jdm}, given by,
\begin{equation}\label{Eq:equilibrium1}
    \begin{split}
        \mathcal{F}_{\nu_e}   & = \mathcal{F}_{\nu_x}    =  \frac{1}{3}(\mathcal{F}^0_{\nu_e}  + 2 \mathcal{F}^0_{\nu_x} ), \\
        \mathcal{F}_{\bar\nu_e}  & =  \mathcal{F}_{\bar\nu_x} = \frac{1}{3}(\mathcal{F}^0_{\bar\nu_e}  + 2 \mathcal{F}^0_{\bar\nu_x} ).\\
    \end{split}
\end{equation}
This phenomenon occurs due to the introduction of non-standard off-diagonal terms in the neutrino-neutrino interaction Hamiltonian, resulting in the emergence of unavoidable neutrino flavor instabilities. These instabilities subsequently trigger flavor decoherence, ultimately leading to a state of flavor equipartition within both the neutrino and antineutrino sectors.

In several beyond Standard Model (BSM) theories of particle physics~\cite{Bialynicka-Birula:1964ddi,Bardin:1970wq}, such neutrino NSSIs are permitted. These interactions are facilitated through a vector mediator, which modifies the effective Lagrangian to $\mathcal{L}_{\mathrm{eff}} \supset G_{\mathrm{F}} [\mathsf{G}^{\alpha\beta} \bar\nu_\alpha \gamma^{\mu} \nu_\beta] [\mathsf{G}^{\xi\eta} \bar\nu_\xi \gamma_{\mu} \nu_\eta]$. This revised Lagrangian closely resembles the neutrino-neutrino interaction Lagrangian of the Standard Model (SM), but now introduces new interaction terms that couple neutrinos of different flavors via $\mathsf{G}^{\alpha\beta}$'s.

\subsection{Magnetically-induced flavor conversion (MFC)}

Neutrinos, which possess exceedingly small yet non-zero magnetic moments (see Refs.~\cite{Giunti:2008ve, Broggini:2012df, Studenikin:2016ykv} for an in-depth review), can exhibit modified flavor evolution when exposed to external magnetic fields. This effect becomes particularly pronounced in environments with ultra-strong magnetic fields (exceeding $10^{15}$ Gauss), such as neutron star mergers  and magneto-rotational CCSNe~\cite{Mosta:2015ucs}. These extreme conditions present ideal scenarios for investigating the interplay between neutrinos and magnetic fields and their impact on collective neutrino oscillations.
For example for Majorana neutrinos, the introduction of a magnetic term can induce neutrino-antineutrino oscillations.

Recent studies  have highlighted the potential significance of the magnetic term when it reaches a comparable magnitude to other terms in the Hamiltonian. 
Here we follow Refs.~\cite{Abbar:2020ggq, Sasaki:2021bvu, Kharlanov:2020cti} that have found an equipartition among $\nu_e$ ($\bar\nu_e$) and $\bar\nu_x$ ($\nu_x$) when the contribution coming from the neutrino magnetic moment to the neutrino Hamiltonian becomes as sizeable as the other terms, described by:
\begin{equation}\label{Eq:equilibrium1}
    \begin{split}
        \mathcal{F}_{\nu_e}   & = \mathcal{F}_{\bar\nu_x}    =  \frac{1}{3}(\mathcal{F}^0_{\nu_e}  + 2 \mathcal{F}^0_{\bar\nu_x} ), \\
        \mathcal{F}_{\bar\nu_e}  & =  \mathcal{F}_{\nu_x} = \frac{1}{3}(\mathcal{F}^0_{\bar\nu_e}  + 2 \mathcal{F}^0_{\nu_x} ).\\
    \end{split}
\end{equation}

\section{Neutrino cross sections}\label{sec:cross}

Before discussing the outcomes of our statistical analysis, let us first briefly outline the neutrino detection channels under consideration in our study.
In our statistical analysis, we consider neutrino
events as can be seen by water Cherenkov detectors, namely Super-Kamiokande and
Hyper-Kamiokande~\cite{Hyper-Kamiokande:2018ofw, Super-Kamiokande:2002weg}.
In particular, we take inverse beta decay (IBD) and neutrino elastic scattering on electrons (ES) as the relevant processes~\footnote{Note that there are a range of other possibilities for neutrino detection which are beyond the scope of this work and we leave them for a future study. 
}.

In this study, we build upon our previous works outlined in Refs.~\cite{GalloRosso:2018ugl, GalloRosso:2017mdz, GalloRosso:2017hbp} 
(see
these References for further details and for the neutrino cross sections).
In particular,  in our calculations, we work under the assumptions of a 100\% detector efficiency and the absence of background events. 
We also ignore the possible existence of a time offset  between 
the start of neutrino emission and
the detection of the first event.
These considerations are made to determine the very minimum number of events necessary to derive conclusive results. 
The energy-differential rate for the reaction $j$ generated
by the neutrino species $\nu_\beta$ can be calculated as,
\begin{equation}\label{eq:rate}
\frac{\mathrm{d} N_{\nu_\beta, j}}{\mathrm{d} E} = n_{\rm{T},j} \int_{E_{\mathrm{th}}}^{\infty}    \mathrm{d} E_\nu\ \mathcal{F_{\nu_\beta}}(E_\nu)  \sigma_{\nu_\beta,j}(E_\nu)
\end{equation}
where $E_{\mathrm{th}}$ is the neutrino energy threshold for each process, $n_{\rm{T},j}$ is the number of targets for the process
$j$, and $\sigma_{\nu_\beta,j}(E_\nu)$ is the energy dependent cross section for a given reaction and
species.

For the IBD events, we use the analytical formula developed in Ref.~\cite{Strumia:2003zx},
\begin{equation}
\sigma_{\rm{IBD}} \simeq 10^{-43}  \mathrm{cm}^2 p_e E_e E_\nu^{x},
\end{equation}
with 
$x = -0.07056 + 0.02018\ \mathrm{ln} E_\nu - 0.001953\ \mathrm{ln}^3 E_\nu$
and
 $E_e \simeq E_\nu - \Delta$, with $\Delta \simeq 1.293$~MeV. For the ES channel, we use tree level expression of
the cross section~\cite{Formaggio:2012cpf,Tomalak:2019ibg}:
\begin{equation}
\frac {\mathrm{d}\sigma_{\nu_\beta, \rm{ES}}}{\mathrm{d} y} = \frac{2 G_{\rm{F}}^2 m_e E_\nu}{\pi} \bigg[  g^2_{\nu_\beta}  
+ g'^2_{\nu_\beta} (1-y)^2 - g_{\nu_\beta} g'_{\nu_\beta} \frac{m_e}{E_\nu} y    \bigg]
\end{equation}
where $g$ and $g'$ can be obtained using the Weinberg weak-mixing angle and are different for different neutrino species (see Table.~3 of Ref.~\cite{GalloRosso:2017mdz}), $G_{\rm{F}}$ is the Fermi constant, and $m_e$ is the electron mass. 
Here, 
 $y$ is the electron kinetic energy, $K_e = E_e - m_e$, divided by the neutrino energy, for which 
$0<y=K_e/E_\nu < (1+m_e/2E_\nu)^{-1}$.

\begin{figure} [tb!]
\centering
\begin{center}
\includegraphics*[width=.45\textwidth, trim= 10 10 10 10, clip]{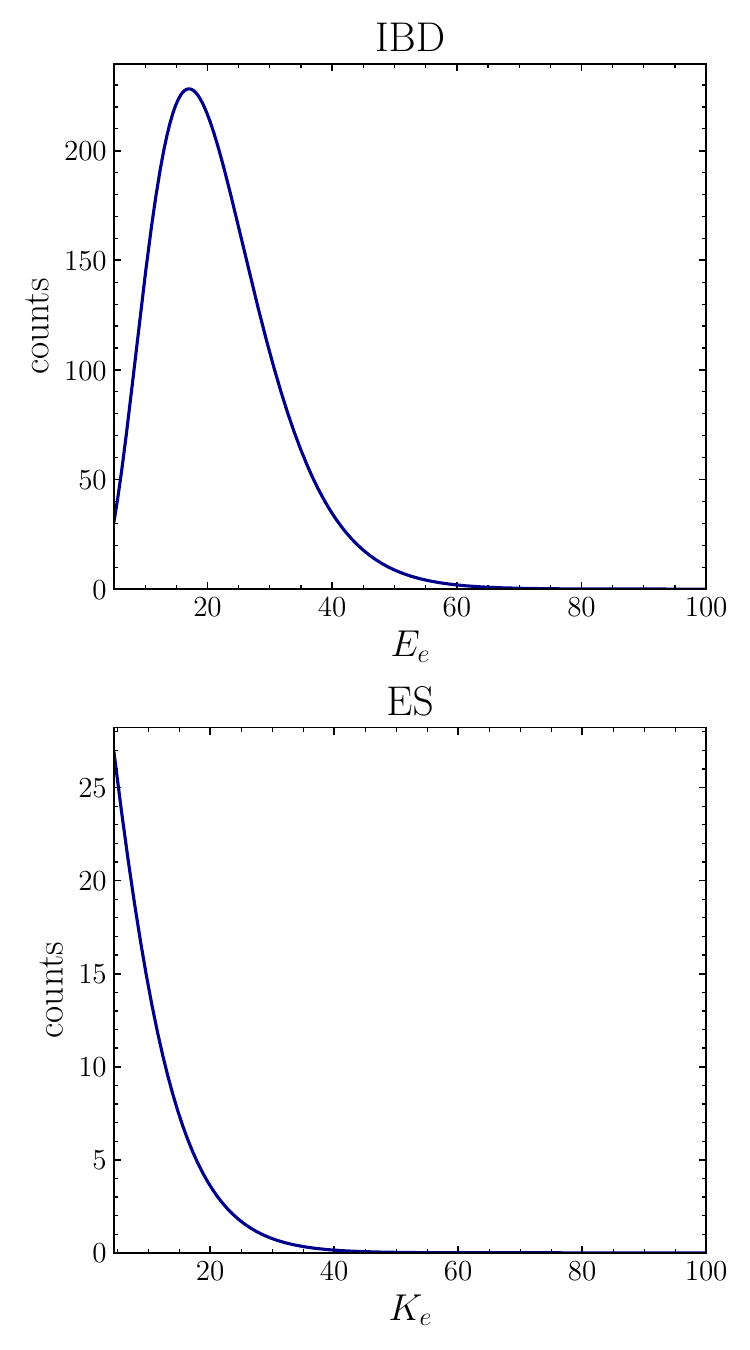}
\end{center}
\caption{
Event rate distributions for the IBD and ES detection channels, as a function of the electron total and
kinetic energies, $E_e$ and $K_e$, respectively. Here, the EQP FC scenario is employed.}
\label{fig:events}
\end{figure}

In Fig.~\ref{fig:events}, we present an illustrative event rate distribution given our equipartition FC prescription, provided both for the IBD and ES detection channels. It is important to note that the anticipated total number of ES events is expected not to exceed a few percent of the IBD events.

\section{Bayesian analysis of the supernova neutrino events}\label{sec:BF}

\begin{figure*} [tb!]
\centering
\begin{center}
\includegraphics*[width=1.\textwidth, trim= 0 0 0 0, clip]{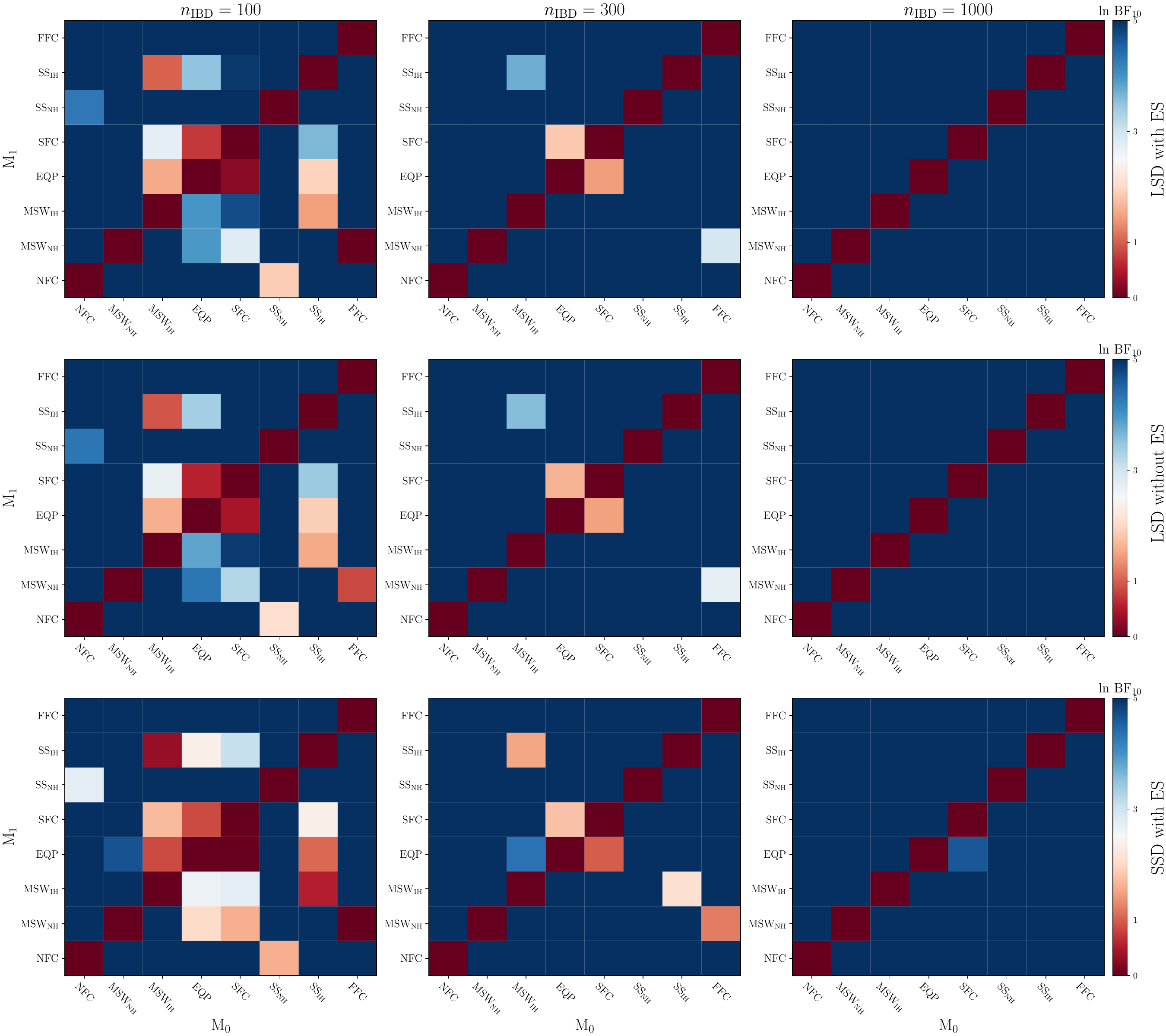}
\end{center}
\caption{
Bayes factor comparison among various  FC scenarios across different observed neutrino events, $N_{\rm{obs}} = 100, 300,$ and 1000. The x-axis denotes the null hypothesis, representing the model by which the  neutrino events were generated, while the y-axis corresponds to the alternative hypothesis. Here, the diagonal values are by default set to zero.
The upper panels show the results using the LSD spectrum, incorporating both the IBD and ES detection channels. In the middle panels, we present the Bayes factor calculations similar to the upper panels, but excluding the ES channel. The lower panels exhibit the Bayes factor values derived from the SSD spectrum calculations (including both IBD and ES channels).}
\label{fig:BF}
\end{figure*}

In this section, we provide the results of our Bayesian analysis to differentiate neutrino FC scenarios via a future SN neutrino signal.
For a specific FC scenario, $\rm{M}_{a}$, an event with an energy $E_i$ has the probability distribution:
\begin{equation}
p(E_i | \mathrm{M}_a) = \frac{1}{\langle N \rangle} \frac{\mathrm{d} N}{\mathrm{d} E},
\end{equation}
where $\mathrm{d} N/\mathrm{d} E$ can be found using Eq.~(\ref{eq:rate}) for each case of FC
 considered, and
$ \langle N \rangle = \int_{E_{\mathrm{th}}}^{\infty} \mathrm{d} E (\mathrm{d} N/\mathrm{d} E)$ is the expected total number of events.

One can then find the  likelihood of observing a set of events with energies $\{E_i \}$ for $i$ in $(1,2,..,N_{\rm{obs}})$ as,
\begin{equation}\label{eq:pr}
P(\{E_i\} | \mathrm{M}_a) = \prod_{i=1}^{N_{\rm{obs}}} p(E_i|\mathrm{M}_a),
\end{equation}
where $N_{\rm{obs}}$ is the observed number of neutrino events.
In our calculation, we derive $N_{\rm{obs}}$ events  using Eq.~(\ref{eq:rate}).
We then calculate the likelihood of observing a set of events with energies $\{E_i\}$ using the above equation.

Assuming that the different FC scenarios have the same priors, one can then find the Bayes factor,
\begin{equation}\label{eq:BF}
 \mathrm{BF}_{10} = \frac{  P(\{E_i\} | \mathrm{M}_1) } { P(\{E_i\} | \mathrm{M}_0)  },
\end{equation}
where $\mathrm{M}_0$ represents the null hypothesis, which is the model the events are generated with, and $\mathrm{M}_1$ is the 
alternative hypothesis to be compared with
 the null hypothesis.  The guidelines for interpreting the natural logarithm of $\mathrm{BF}_{10}$ are as follows:
i) $\mathrm{ln\ BF}_{10} \in (0–1)$: Not significant, 
ii) $\mathrm{ln\ BF}_{10} \in (1–3)$: Indicates a positive evidence,
iii) $\mathrm{ln\ BF}_{10} \in (3–5)$: Reflects strong evidence, and
iv) $\mathrm{ln\ BF}_{10}  > 5$: Demonstrates very strong evidence.

In line with Ref.~\cite{Hyper-Kamiokande:2021frf}, our analysis revolves around a fixed number of  observed neutrino events, irrespective of the absolute SN neutrino luminosities  and its distance from the detector. 
In this context, the absolute neutrino luminosities of the  SN and its distance from the detector become irrelevant in our evaluations, i.e., they can not be considered as free parameters. This is because, in any case, specific values for these parameters must be assumed to obtain the given number of events for each  FC scenario under consideration.
What is important in our analysis  are the spectral parameters and the  relationship between the luminosities of various neutrino species. 

We consider calculations in which   the number of observed events are 100, 300, and 1000.
In addition, in our calculations, we assume a constant ratio between the number
of observed events in the IBD and ES channels, namely $N_{\rm{ES}} = 0.05\ N_{\rm{IBD}}$, consistent with the expectations 
in a water Cherenkov detector.

\subsection{Analysis assuming known spectral parameters}

In our first evaluation of the differentiating between various FC scenarios, we work under the assumption that the spectral parameters are known, essential for calculating the likelihoods outlined in Eqs.~(\ref{eq:pr}) and (\ref{eq:BF}). This assumption stands on two  grounds. Firstly, it offers an initial estimation of our capability of differentiating between various FC scenarios. Should the FC models remain indistinguishable given the spectral parameters, discerning them in situations requiring optimisation for obtaining  spectral parameters (as discussed in the following section) becomes a considerable challenge. 
Moreover, it is plausible to consider the availability of such spectral information, particularly in models tied to specific progenitors 
once multi-dimensional simulations by different groups will reach convergence.
Nevertheless, these preliminary calculations yield crucial initial insights that prove invaluable later in the analysis.

In Fig.~\ref{fig:BF}, we present the Bayes factor obtained from Eq.(\ref{eq:BF}) across various FC scenarios, given the spectral parameters. The results are shown for $N_{\rm{obs}} = 100, 300,$ and 1000. The upper panels illustrate outcomes derived from events generated using the LSD spectrum, with more pronounced differences among neutrino spectra, where we have considered both the IBD and ES detection channels.  
Our approach involves running 1000 Monte Carlo (MC) simulations for each scenario, subsequently averaging the results to obtain the Bayes factor's average. 

Let us make
a few remarks. Firstly, even with $N_{\rm{obs}} = 100$, discerning certain FC models becomes feasible. However, a larger event count significantly enhances our ability to differentiate between the FC scenarios. Then upon reaching $N_{\rm{obs}} = 1000$, perfect differentiation among all the FC models becomes achievable.
It is also illuminating to note  that in these calculations, we have not consider the MFC  scenario, since $\nu_x$ and $\bar\nu_x$ are indistinguishable observationally,
 MFC behaves closely to that of the EQP scenario.

To provide the reader with some insight into the Bayes factor values, Table~\ref{table:BF} presents $\mathrm{ln\ BF}_{10}$ corresponding to the LSD calculation with $N{\rm{obs}} = 300$. It is important to note that several computations exhibit clear distinguishability among FC scenarios. However, certain entries in the table display values below 10, suggesting a potential lack of strong differentiation when considering the whole range of  MC calculations. In specific cases, the samples with the most adverse impact on the Bayes factor might reduce  $\mathrm{ln\ BF}_{10}$ value to less than 5.

\begin{table} [tb!]
\centering
\begin{center}
\includegraphics*[width=.5\textwidth, trim= 0 0 0 0, clip]{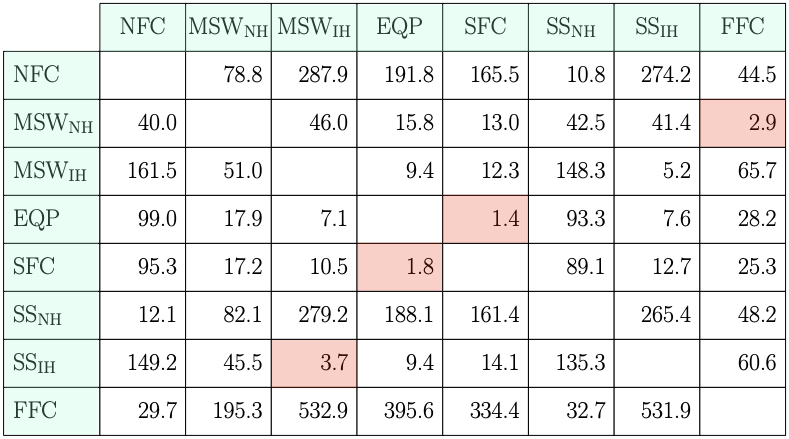}
\end{center}
\caption{
 $\mathrm{ln\ BF}_{10}$ values corresponding to the LSD calculation with $N_{\rm{obs}} = 300$ (the upper middle panel of Fig.~\ref{fig:BF}). 
 The rows/columns indicate the null/alternative hypothesises. 
 The red cells show the values below 5.
Note that several cells show clear distinguishability among FC models. However, certain entries in the table display values below 10, suggesting a potential lack of strong differentiation when considering the whole range of  MC calculations.}
\label{table:BF}
\end{table}

In the middle panels of Fig.~\ref{fig:BF}, we display the Bayes factor for the calculations closely resembling those in the upper panel with the LSD spectrum, except for the exclusion of the ES channel. It is important to note the minimal discrepancy between the upper and middle panels, indicating that the primary influential factor here is  the number of observed neutrino events. The ES channel, while included in the calculations of the upper panel, appears to play a subdominant role and does not significantly impact the outcomes.

In the lower panels of Fig.~\ref{fig:BF}, the Bayes factor for the SSD spectrum calculations is displayed, revealing more subtle variations across the neutrino spectra. Comparatively, the distribution of the Bayes factor appears less prominent than in the middle and upper panels. However, large values of $\mathrm{ln\ BF}_{10}$ can still be observed and most of the FC scenarios can be distinguished well from each other. This suggests that even in less favorable scenarios, there remains strong potential for discerning between different FC scenarios.

\subsection{FC scenarios, including the MSW effect}

\begin{figure*} [tbh!]
\centering
\begin{center}
\includegraphics*[width=.95\textwidth, trim= 0 0 0 0, clip]{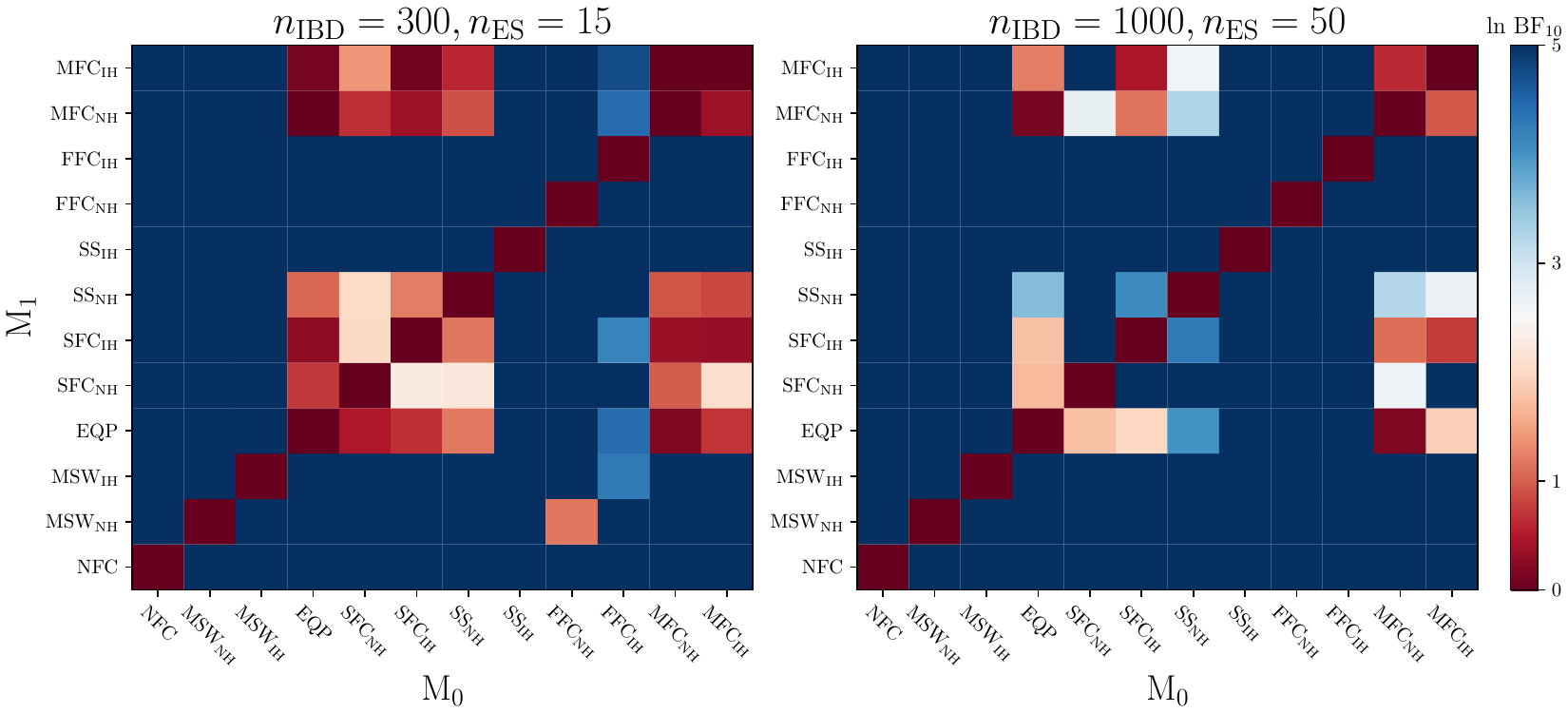}
\end{center}
\caption{
The Bayes factor regarding  FC scenarios with MSW. Note that here we have considered the LSD spectrum and known spectral parameters.
The x-axis denotes the null hypothesis, representing the model by which the  neutrino events were generated, while the y-axis corresponds to the alternative hypothesis. The diagonal values are by default set to zero.}
\label{fig:MO}
\end{figure*}

In the preceding part, we focused on a single FC scenario. However and in practice,  
neutrinos are expected to undergo the MSW effect after encountering FC mechanisms in the deeper SN zones.
Hence, it follows that each FC scenario should encompass an MSW effect on top, either within the NH or IH paradigm. 
In our computations, we consider various FC scenarios, and the final neutrino fluxes which are detailed in Sec.\ref{sec:FC}. Given that the MSW effect is expected to occur far from other FC scenarios, we proceed by assuming that the MSW effect can be superimposed. In this context, we consider the final $\mathcal{F}_\nu$ post FC scenario as an input ($\mathcal{F}^0_\nu$) to Eqs.(\ref{eq:nh}) and (\ref{eq:ih}), accounting for the MSW effect.

In Fig.~\ref{fig:MO}, we illustrate the Bayes factor regarding  FC scenarios with MSW. In this context, we have considered the LSD spectrum assuming known spectral parameters, as above, and with $N_{\rm{obs}} = 300$ and $1000$ . It is evident that the introduction of FC scenarios with MSW on top amplifies the degeneracy, thereby complicating the distinction between different FC scenarios. Note that the EQP scenario does not include any MSW on top, since the existing equipartition is already blind to MSW effect.

\subsection{Unknown spectral parameters}

\begin{figure*} [tb!]
\centering
\begin{center}
\includegraphics*[width=1.02\textwidth, trim= 0 0 0 0, clip]{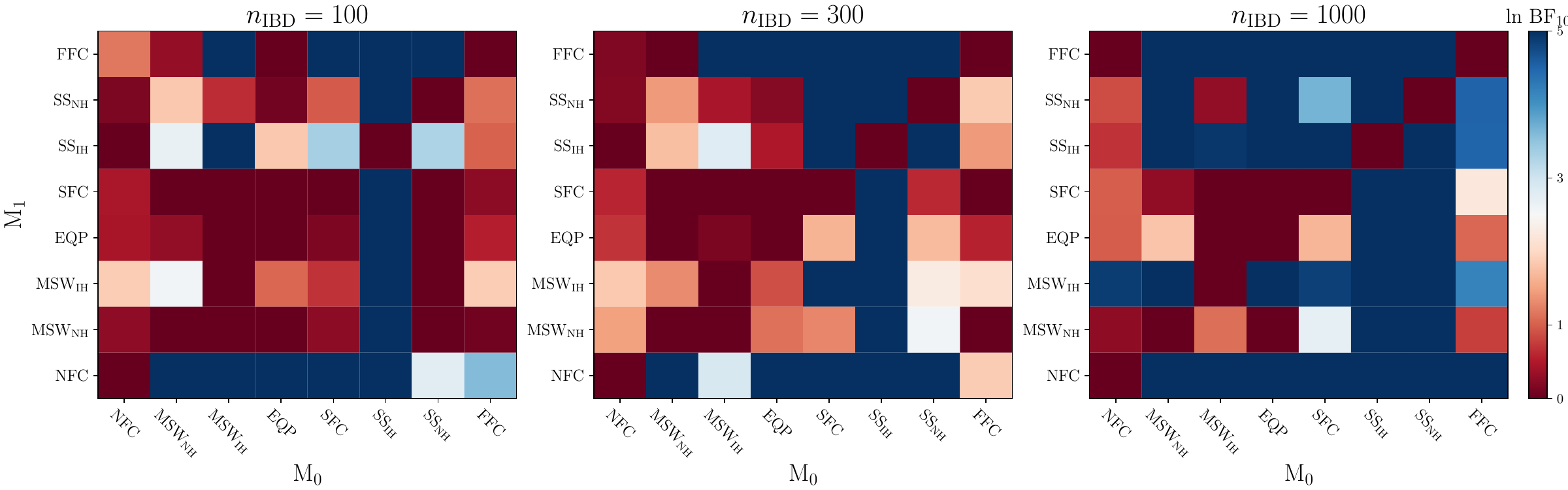}
\end{center}
\caption{
The Bayes factor regarding the case where the spectral parameters are not known beforehand. 
The x-axis denotes the null hypothesis (obtained from the LSD spectrum), representing the model by which the  neutrino events were generated, while the y-axis corresponds to the alternative hypothesis. The diagonal values are by default set to zero.
}
\label{fig:MC}
\end{figure*}

\begin{figure} [tb!]
\centering
\begin{center}
\includegraphics*[width=.5\textwidth, trim= 0 0 0 0, clip]{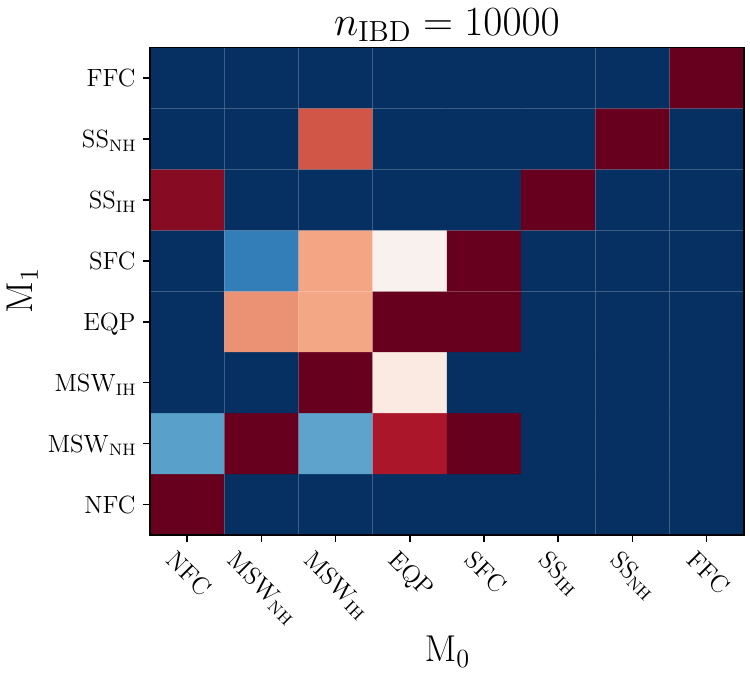}
\end{center}
\caption{
The Bayes factor regarding the case where the spectral parameters are  unknown, with $n_{\rm{IBD}}=10000$.
The x-axis/y-axis denote the null/alternative hypothesises.
}
\label{fig:MC_2}
\end{figure}

Up to this point, our approach involved assuming that the spectral parameters were known once the Bayes factor was calculated. However, in reality, the Bayes factor remains unknown at the time of detection. As a result, one should first optimize over the likelihood distribution to determine the most probable spectral parameters. In Fig.~\ref{fig:MC}, we show our findings in such a scenario.
In our computations, we initially optimized the likelihood for each model, obtaining the optimal spectral parameters and subsequently, we calculated the likelihood for each FC scenario. 
During  the optimization process, we assume the spectral parameters to be confined in the following intervals: 
\begin{equation}
\begin{split}
&L_{\bar\nu_e}/L_{\nu_e},\ L_{\nu_x}/L_{\nu_e} \in (0.1, 1.5),\\ 
&\langle E_{\nu_e} \rangle \in (5,15),\ \langle E_{\bar\nu_e} \rangle \in (8,18),\ \langle E_{\nu_x} \rangle \in (10,20)\ \mathrm{MeV}, \\
&\alpha_{\nu_e} \in (2.5,4),\ \alpha_{\bar\nu_e} \in (3.5,5),\ \alpha_{\nu_x} \in (1.5,3),\\ 
\end{split}
\end{equation}
which could be considered conservative enough for the SN accretion phase. 
To generate the neutrino signal for each model, we utilized the LSD spectrum in these calculations. In addition, being guided by the results in the previous section, we only consider the IBD detection channel to enhance the computation efficiency.

When spectral parameters remain unknown, distinguishing between various FC scenarios becomes notably challenging, necessitating a larger number of observed neutrino events. However and despite this complexity, there remains optimism regarding the possibility of distinguishing between different FC scenarios. 

In our current study, we have focused on $N_{\rm{obs}}$ values up to 1000. However, it is plausible to extend this analysis to encompass a larger number of neutrino events during the SN accretion phase, especially when incorporating the Hyper-Kamiokande or/and  the inclusion of multiple detectors is considered. In Fig.~\ref{fig:MC_2}, we present calculations where $N_{\rm{obs}} = 10000$. This expanded dataset distinctly separates a larger number of  FC scenarios, showing their discernibility from one another.

\section{DISCUSSION AND OUTLOOK}\label{sec:dis}

Neutrino FC within the  SN environment is an intricate and multifaceted phenomenon. Previous studies have shown that varying physics within the SNe can yield diverse FC scenarios affecting neutrino flavor evolution  in these extreme settings.
In this study we have employed  Bayesian techniques to discern and differentiate between various FC scenarios in the SN environment using the neutrino signal   from a future galactic CCSN. Our findings demonstrate a promising prospect: with a sufficiently large number of observed events (exceeding a few hundred), there exists a high probability of distinguishing between different FC scenarios based on the neutrino signals emitted by the next galactic CCSN.

In our analysis, we examine two distinct neutrino energy spectra designed to characterize the neutrino spectra throughout the SN accretion phase. One spectrum exhibits more pronounced variations among neutrino species, while the other displays comparatively minor fluctuations.
Our findings illustrate that while distinguishing among the  FC scenarios may be more discernible in the former spectrum, there remains a considerable probability of distinguishing between different FC scenarios even in cases where spectral differences are less pronounced.

In addition, in our calculations, we incorporate two distinct neutrino detection channels: IBD and ES. While the ES channel exhibits sensitivity to all neutrino species and possesses the potential to distinguish among different neutrino flavors, our findings illustrate that the crucial factor here lies in the events number. Surprisingly, our results indicate that incorporating the ES channel does not significantly enhance our analysis due to its relatively lower number of events.

In realistic scenarios it is crucial to consider two important factors. The first is that neutrinos may encounter FC scenarios with MSW effect on top. This means they could undergo flavor conversions deep within the  SN core while also experiencing the MSW effect at larger radii. Moreover, the spectral parameters of neutrinos might remain unknown at the time of detection, which necessitates an optimization to finding the spectral parameters. 
Our findings demonstrate that both these effects unfavorably impact our ability to distinguish among different FC scenarios and increase the minimum number of events required for efficient FC scenario differentiation.

In summary, our findings highlight the potential to differentiate various FC scenarios within the  SN environment by analyzing a future SN neutrino signal. However, there remain crucial paths for further investigation.
Primarily, our calculations excluded any  information regarding the SN models. 
One can, in principle, link various  FC scenarios to  SN models derived from CCSN simulations. This connection has been explored to some extent in the studies detailed in Refs.~\cite{Hyper-Kamiokande:2021frf,Olsen:2022pkn,Saez:2024ayk}, though only focusing  on the  MSW effect. The incorporation of SN models has the potential to improve the ability to differentiate between FC scenarios in calculations where spectral parameters are unknown. However, it is crucial to note that introducing the complexity of various SN models might simultaneously hinder the distinguishability of different FC scenarios by introducing a potential degeneracy.
Additionally, our FC scenarios presumed perfect conversions, which may not hold true in reality. For example, the MSW effect might not achieve complete adiabaticity, resulting in final flux  values deviating from those of an ideal adiabatic MSW scenario. Similar uncertainties could exist in other FC scenarios, where actual equipartition might only be partial, differing from our assumed perfect conditions. Assessing the distinguishability of various FC scenarios under partial conversions would be a valuable exploration.
Furthermore, our focus was solely on the IBD and ES detection channels. While our results suggest that incorporating the ES channel does not significantly impact our analysis, this might differ for other available neutrino detection channels.
Considering the imperative need to comprehend neutrino FC mechanisms within the SN environment, addressing these crucial aspects significantly enhance our ability to extract insights from the neutrino signal of a future galactic CCSN.


\section*{Acknowledgments}
S.A. is deeply grateful to Georg Raffelt for insightful conversations.
S.A. was supported by the German Research Foundation (DFG) through
the Collaborative Research Centre  ``Neutrinos and Dark Matter in Astro-
and Particle Physics (NDM),'' Grant SFB-1258, and under Germany’s
Excellence Strategy through the Cluster of Excellence ORIGINS
EXC-2094-390783311.
We would also like to acknowledge the use of the following softwares: \textsc{Scikit-learn}~\cite{pedregosa2011scikit}, 
\textsc{Keras}~\cite{chollet2015keras},
\textsc{Matplotlib}~\cite{Matplotlib}, \textsc{Numpy}~\cite{Numpy}, \textsc{SciPy}~\cite{SciPy}, and \textsc{IPython}~\cite{IPython}.

\bibliographystyle{elsarticle-num}
\bibliography{Biblio}

\clearpage

\end{document}